\documentstyle[12pt]{article}
\newcommand \be {\begin{equation}}
\newcommand \bea {\begin{eqnarray} \nonumber }
\newcommand \ee {\end{equation}}
\newcommand \eea {\end{eqnarray}}
 \newcommand \eps {\epsilon}
 \newcommand \s {\sigma}

 \newcommand \g {\gamma}
\newcommand \la {\lambda}

 \newcommand \al {\alpha}

 \newcommand \ba {\overline}
\newcommand \lan {\langle}
 \newcommand \ran {\rangle}
\newcommand \bi {\bibitem}
\newcommand \Tr {\mbox{Tr}}

\catcode`\@=11
\long\def\@makefntext#1{
\protect\noindent \hbox to 3.2pt {\hskip-.9pt
$^{{\ninerm\@thefnmark}}$\hfil}#1\hfill}                

 \def\@makefnmark{\hbox to 0pt{$^{\@thefnmark}$\hss}}  

\def\ps@myheadings{\let\@mkboth\@gobbletwo
\def\@oddhead{\hbox{}
\rightmark\hfil\ninerm\thepage}
\def\@oddfoot{}\def\@evenhead{\ninerm\thepage\hfil
\leftmark\hbox{}}\def\@evenfoot{}
\def\sectionmark##1{}\def\subsectionmark##1{}}

\newcounter{sectionc}\newcounter{subsectionc}\newcounter{subsubsectionc}
\renewcommand{\section}[1] {\vspace{0.6cm}\addtocounter{sectionc}{1}
\setcounter{subsectionc}{0}\setcounter{subsubsectionc}{0}\noindent
        {\bf\thesectionc. #1}\par\vspace{0.4cm}}
\renewcommand{\subsection}[1] {\vspace{0.6cm}\addtocounter{subsectionc}{1}
        \setcounter{subsubsectionc}{0}\noindent
        {\it\thesectionc.\thesubsectionc. #1}\par\vspace{0.4cm}}
\renewcommand{\subsubsection}[1]
{\vspace{0.6cm}\addtocounter{subsubsectionc}{1}
        \noindent {\rm\thesectionc.\thesubsectionc.\thesubsubsectionc.
        #1}\par\vspace{0.4cm}}

\newcounter{appendixc}
\newcounter{subappendixc}[appendixc]
\newcounter{subsubappendixc}[subappendixc]

\renewcommand{\appendix}[1] {\vspace{0.6cm}
        \refstepcounter{appendixc}
        \setcounter{figure}{0}
        \setcounter{table}{0}
        \setcounter{equation}{0}
        \renewcommand{\thefigure}{\Alph{appendixc}.\arabic{figure}}
        \renewcommand{\thetable}{\Alph{appendixc}.\arabic{table}}
        \renewcommand{\theappendixc}{\Alph{appendixc}}
        \renewcommand{\theequation}{\Alph{appendixc}.\arabic{equation}}
        \noindent{\bf Appendix \theappendixc #1}\par\vspace{0.4cm}}

\def\abstracts#1{{
        \centering{\begin{minipage}{30pc}\tenrm\baselineskip=12pt\noindent
        \centerline{\tenrm ABSTRACT}\vspace{0.3cm}
        \parindent=0pt #1
        \end{minipage}}\par}}

\renewenvironment{thebibliography}[1]
        {\begin{list}{\arabic{enumi}.}
        {\usecounter{enumi}\setlength{\parsep}{0pt}
\setlength{\leftmargin 1.25cm}{\rightmargin 0pt}
         \setlength{\itemsep}{0pt} \settowidth
        {\labelwidth}{#1.}\sloppy}}{\end{list}}

\newcounter{itemlistc}
\newcounter{romanlistc}
\newcounter{alphlistc}
\newcounter{arabiclistc}

\newcommand{\fcaption}[1]{
        \refstepcounter{figure}
        \setbox\@tempboxa = \hbox{\tenrm Fig.~\thefigure. #1}
        \ifdim \wd\@tempboxa > 6in
           {\begin{center}
        \parbox{6in}{\tenrm\baselineskip=12pt Fig.~\thefigure. #1}
            \end{center}}
        \else
             {\begin{center}
             {\tenrm Fig.~\thefigure. #1}
              \end{center}}
        \fi}

\newcommand{\tcaption}[1]{
        \refstepcounter{table}
        \setbox\@tempboxa = \hbox{\tenrm Table~\thetable. #1}
        \ifdim \wd\@tempboxa > 6in
           {\begin{center}
        \parbox{6in}{\tenrm\baselineskip=12pt Table~\thetable. #1}
            \end{center}}
        \else
             {\begin{center}
             {\tenrm Table~\thetable. #1}
              \end{center}}
        \fi}

\def\@citex[#1]#2{\if@filesw\immediate\write\@auxout
        {\string\citation{#2}}\fi
\def\@citea{}\@cite{\@for\@citeb:=#2\do
        {\@citea\def\@citea{,}\@ifundefined
        {b@\@citeb}{{\bf ?}\@warning
        {Citation `\@citeb' on page \thepage \space undefined}}
        {\csname b@\@citeb\endcsname}}}{#1}}

\newif\if@cghi
\def\cite{\@cghitrue\@ifnextchar [{\@tempswatrue
        \@citex}{\@tempswafalse\@citex[]}}
\def\citelow{\@cghifalse\@ifnextchar [{\@tempswatrue
        \@citex}{\@tempswafalse\@citex[]}}
\def\@cite#1#2{{$\null^{#1}$\if@tempswa\typeout
        {IJCGA warning: optional citation argument
        ignored: `#2'} \fi}}

\def\fnt#1#2{\footnotetext{\kern-.3em
        {$^{\mbox{\sevenrm #1}}$}{#2}}}

 1
 1
 1

\font\tenbf=cmbx10
\font\tenrm=cmr10
\font\tenit=cmti10

\font\ninerm=cmr9

\begin{document}

\centerline{\tenbf GAUGE THEORIES, SPIN GLASSES AND REAL GLASSES}

\baselineskip=16pt 
\centerline{\ninerm Talk presented at the Oskar Klein Centennial Symposium}

\baselineskip=22pt
\vspace{0.8cm}
\centerline{\tenrm GIORGIO PARISI}
\baselineskip=13pt
\centerline{\tenit Dipartimento di Fisica,
Universit\`a  La  Sapienza and  INFN Sezione di Roma I}
\baselineskip=12pt
\centerline{\tenit Piazzale
Aldo Moro, Roma 00187, Italy}
\vspace{0.9cm}

\abstracts{In this talk I will show that  usual spin glasses are
a peculiar kind of Abelian gauge
theory. I will shortly review the techniques used to study them.
At the end I will consider more
general models (e.g. spin glasses based on non Abelian gauge group) and I will
discuss  the relevance of these models to
real glasses. Finally I will derive from first  principles a generalised
Vogel-Fulcher law for the divergence of the characteristic time near the glass
transition. }

\vfil
\rm\baselineskip=14pt
\section{Introduction}

The aim of my talk is to show how some of the subjects on which Oscar
Klein worked (in particular I
refer to gauge theories) have been useful in  apparently far away
subjects, i.e. spin glasses and
real  glasses. Indeed some models of spin glasses are a gauge
theories of a rather peculiar type.

I will start (in the second section) by briefly recalling the
definition of a spin glass and presenting an
idealised model. Later (in the third section) I will show that this
model is in reality a gauge theory
corresponding to the $Z_2$ group. Next (in the forth section) I will
review the basic ingredients of
the  replica method which are used to study these models. I will later
(in the fifth section) consider
Abelian and  non Abelian generalisations and I will derive from first
 principles a generalised
Vogel-Fulcher law for the divergence of the characteristic time near
the glass transition.
In the last section I will
argue that these non Abelian
models may share   many properties with real glasses.
\section{Spin Glasses}

I  introduce  here a very simple model of spin glasses\cite{EA,Bi86}. I
consider
a material where there are three  kinds of atoms: $M$,  $A$  and  $B$.  $M$  is
a
magnetic atom (it has  a non zero magnetic moment)  while  $A$ and $B$ are
magnetically inert.

I suppose that at low temperature the system crystallises in such a  way that
the
$M$-atoms stay on a  regular lattice and the $A$ and $B$ atoms stay on the
links of
the same lattice. The position of the  magnetically inert atoms is supposed to
be
random. In other words we  consider an $M\ A_x \  B_{100-x}$ alloy; the case
$x=50$
corresponds to an equal proportion of $A$  and $B$.

Let us assume that the magnetic interaction among the $M$-atoms is  of the
nearest
neighbour type and that it is mediated  by the non magnetic atoms. The
interaction among two
$M$-atoms is  ferromagnetic if the link is  occupied by an $A$ atom, while it
is
antiferromagnetic if the link is  occupied by a $B$ atom. We also  assume for
simplicity that the strengths of the ferromagnetic and of  the
antiferromagnetic
interactions are equal.

Usually the magnetic interaction is relevant only at temperatures  much lower
than
the melting  temperature and it may be neglected during the formation of the
alloy.
At low temperature the atoms  may only oscillate around their equilibrium
positions.
If the  temperature is decreased fast  enough the position of the atoms in not
influenced by the magnetic  interaction. We can describe this situation by
saying
that we are in presence of a {\sl quenched}  disorder.

If we assume that the spin are Ising variables, the corresponding  Hamiltonian,
in
presence of a  magnetic field $h,$ is
\be
H_U[\s] \equiv - \sum_{i,k}\s_iU_{i,k}\s_k  -h \sum{\s_i}\ . \label{HAMIL}
\ee
The variables $\s$ are defined on the sites of the lattice and they take
the values $\pm 1$. The variables
$U_{i,k}$ are defined on the links of the lattice, i.e. when $i$ and
$k$ are nearest neighbours;
they also take the values $\pm 1$.

The variables $U$ are random independent variables. For each choice
of the $U$ we can define a
statistical expectation value:
\be
\lan g(\s) \ran_U ={\sum_{\s}\exp(-\beta H_U[\s]) g(\s) \over
\sum_{\s}\exp(-\beta H_U[\s])  }.
\label{QUENCH}
\ee
We are interested in computing the statistical expectation values
averaged over the probability distribution of the samples, in other
words the quantity
\be
\lan g(\s) \ran \equiv \ba {\lan g(\s) \ran _U}
\equiv \int dP(U) \lan g(\s) \ran_U \ ,\label {AVE}
\ee
where we denote by an horizontal bar the average over the $U$ and $P(U)$
is the probability distribution of the variables $U$. We
have already remarked that the
probability distribution of the variables $U$ does not depend from
their interaction with the $\s$:
for each physical realization of the system they  are fixed
(quenched) variables.

 In the infinite volume limit the expectation value of
intensive quantities does not depend on the  realisation of the couplings $U$
 (i.e. all the samples have essentially the
same properties) and the sample to sample  fluctuations vanish in
this limit.

A very interesting quantity to evaluate is the magnetic
susceptibility. A naive computation would
give the following formula
 \be
\chi=\beta(1- \ba{ \lan \s_i \ran_U ^2}) = \beta(1- \ba{ m(i)_U ^2}) = \beta
(1-
q_{EA})\ ,
\ee
where $m(i)_U\equiv \lan \s_i \ran_U$ is the site dependent spontaneous
magnetisation and $q_{EA} \equiv
\ba{ m(i)_U ^2}$ is  the so called Edward Anderson order parameter\cite{EA}.
We shall see later how this
formula for the susceptibility is modified by a more sophisticated treatment.

Physical intuition tell us
that at high temperature at zero magnetic field there is no
spontaneous
magnetization and consequently
$q_{EA}=0$. At low temperature each sample should develop its own
spontaneous magnetization, and
consequently $q_{EA}\ne 0$ at low temperature. The non vanishing of
$q_{EA}$ should therefore
mark the spin glass transition.

Before studying this model further it is convenient to analise its
symmetries and in particular the
consequences of gauge  invariance.

\section{Gauge invariance}

Let us consider two different systems such that
 the couplings of the two systems are equal in all links not  connected to a
given
point $i$ and they differ by a sign only for  the links connected to that
point.
These two sets of couplings  have
the same probability to be realised (all sets of couplings have the same
probability). They have also the same free energy
because  the Hamiltonian does not change (at zero  magnetic
field) if simultaneously we  change sign to the  spin
variable sitting at the point.

In other words both the measures (on $U$ and $\s$) and the
Hamiltonian are invariant with respect to
the local gauge transformation\cite{TOU}:
\be
\s_i \to -\s_i \ ;\ \ \ \ \ \ \ \ \ U_{i,k} \to U_{i,k}\ .
\ee

The set of all possible realizations of the system  at zero magnetic  field is
 gauge
invariant under the gauge group $Z_2$.  The couplings  and the spins play
respectively the role of the gauge connection and of the matter field. At
non-zero
magnetic field  the gauge invariance is  explicitly broken. The Hamiltonian at
 zero
magnetic field is  (apart from a constant) the  square of the covariant
lattice-gradient of the $\s$ variables. It can be written as
\be
\sum_i \sum_\mu (\s_i-U_{i,i+\mu} \s_{i+\mu})^2\ .
\ee

The relevant quantities  are  gauge invariant. If
two realizations of the couplings $U$ and $U'$ differ by a gauge
transformation, their thermodynamic
properties are the same. It is important to concentrate our attention
on gauge invariant quantities.

Let us study in more details how the thermodynamical quantities depend  on the
 choice
of the couplings. A quantity which is often used to characterise the gauge
fields is
the  Wilson loop. We can associate to each closed circuit on the lattice the
ordered product  of all the links of the circuit.
We thus define:
\be
W(C)  \equiv \prod_{(i,k) \in C} U_{i,k}\ .
\ee
This quantity may take the values $\pm 1$. If $W(C)=-1$, the loop is
said to be {\sl frustrated}\cite {TOU}: along that loop it
is not possible to find a configuration of spins  such
that
\be
U_{i,k}\s_i \s_k = 1 \ \ \ \forall \ (i,k) \in C.
\ee
If there are frustrated loops (i.e. if there is no gauge transformation
which brings all couplings $U$ to 1) it is not
possible to find a configuration of the spins such that all terms in
the Hamiltonian are positive.
Some defects (i.e. links for which the contribution to the energy is
negative) must be present.

At low temperature the equilibrium probability distribution is
concentrated on those
spin configurations which have the minimal energy. It is interesting
to study also those
configurations which are local minima of the Hamiltonian in the sense
that the Hamiltonian increases
when we flip a spin. These local minima are very important in the
dynamics outside equilibrium
because at low temperatures the system may be trapped for a very long
time in these minima.

If we study the structure of local and global minima with
great care, we  discover that
frustration implies the presence of defects which can be put in many  ways on
the
lattice. The ground state is degenerate. The number of local and not global
minima
is  also large.

This phenomenon is well known in gauge field. On the lattice the
choice of the Landau gauge
corresponds to find the maximum of
\be
\sum_{i,k} \Tr g^\star_i U_{i,k} g_k,
\ee
where $g$ is the gauge transform which brings the gauge fields
in the Landau gauge\cite{MPRI}. This problem is equivalent (in the
our case) to find the minimum of
the Hamiltonian eq. (\ref{HAMIL}).  Gribov ambiguity tell us that in the
general case there are many possibility of
choosing the gauge. This result implies the existence of many minima of
the
Hamiltonian eq. (\ref{HAMIL})\footnote{In the continuum Gribov ambiguity
is normally present in non  Abelian gauge theories. It is also
present in Abelian theories when we allow configurations which are
singular in the continuum limit, e.g.
like magnetic monopoles.}.

 \section{The Replica Method}

We face now the problem of evaluating
 the quantities which appear in equation  (\ref{QUENCH}).
 The
first proposal would be to sum over the variables $U$ and to remain with an
effective interaction for the  variables $\s$. This approach clashes with fact
 that
both the numerator and the denominator of eq. (\ref{QUENCH}) depend on the
 variables
$U$  and the sum over the $U$ is not easy.

This difficulty may be avoided by introducing $n$ identical copies
(or replicas) of the same system \cite {EA}.
We define \be
\lan g(\s) \ran_n ={\sum_{\s}\sum_U\exp(-\beta
 H_n[\s,U]) g(\s^1) \over
                    \sum_{\s}\sum_U\exp(-\beta H_n[\s,U]) },
\ee
where the spins $\s^a_i$ carry an other index ($a$) which ranges from one to
$n$.
The new Hamiltonian is the sum of
$n$ identical Hamiltonians
\be
 H_n(\s,U)= \sum_{a=1,n} H_U[\s^a].
\ee

It easy to check that
\be
\lan g(\s) \ran_n= {\sum_U \lan g(\s) \ran_U (Z_U)^n \over
                    \sum_U  (Z_U)^n },
\ee
where the $U$-dependent partition function is defined as
\be
 Z_U=\sum_{\s}\exp(-\beta H_U[\s]).
\ee
We finally find that
\be
\lan g(\s) \ran =\lan g(\s) \ran_n \vert _{n=0} \ .
\ee

In this way one finds that properties of the matter fields averaged
over the disordered
(quenched) gauge fields can be computed by considering the gauge
fields interacting with $n$ copies
of the matter field and computing the expectation values in the limit
$n \to 0$. The argument is familiar to those who work in numerical
simulation of latttice gauge theories, where $n$ plays the role of the
number of quark flavours in the sea\footnote{It is evident  why I
proposed the name {\sl  quenched} for the approximation of neglecting
quark loops in QCD.}.

The average of the $U$ fields can be done and one remains with an
effective interaction for the $\s$
variables. This effective interaction must be written in terms of the
gauge invariant combinations.
In this case the most appropriate variables are
\be
Q^{ab}_i=\s^a_i \s^b_i.
\ee
The diagonal terms of the matrix $Q$ are identically equal to 1, so
that we can consider only the
off-diagonal term. All computations must be done for generic
values of $n$ and
we must send $n$ to zero at
the end.

As usually we can expand the effective interaction in powers of $Q$,
being careful to preserve the various symmetries of
the problem, i.e.:
\begin{itemize}
\item
 The  group of permutations of the $n$ replicas $S(n)$.
 \item The spin reversal symmetry for each replica, i.e. $n$ times
the direct product of the $Z_2$ global group, where each
$Z_2$ group
acts on a different replica\footnote{This symmetry is present only
at zero magnetic field.}.
\end{itemize}

In the continuum limit
in $D$ dimensions the simplest form for the effective free energy, in which all
important terms are  contained, is the following :
 \be
F[Q]=\int d^Dx \left({1 \over 2}\sum_\mu \Tr(\partial_\mu
Q(x))^2+W(Q(x))\right),
\ee
where the function $W(Q)$is given by
\be
W(Q)=\tau \Tr(Q^2) +g_3 \Tr(Q^3) + g_4\Tr (Q^4) +y \sum_{ab}Q^4_{ab} \ .
\ee

The usual strategy consists in computing the minimum of $F$ and
constructing the usual perturbative
expansion for the small fluctuations around this minimum (the so
called loop expansion). As we shall see even the first step is not
very simple.

Let us assume that the $F[Q]$ is minimized by a function $Q(x)$ which
does not depend from $x$. We can this set $Q(x)=Q$ and look for the
minimum of $W(Q)$.

For $\tau>0$ we find that there is only a minimum at $Q=0$. When $\tau$ is
negative we  easily find that there is a stationary points at
\be
Q_{ab}=q,\ \ \     \forall \ a \ b.\label{SYM}
\ee
This point is invariant under the replica group $S(n)$. In order to
decide if this stationary point is a
minimum of $W$, we have to compute the small fluctuations around this
point. The Hessian
\be
{\cal H}_{ab,cd} ={\partial W \over \partial Q_{ab} Q_{cd}}\ ,
\ee
must have non negative eigenvalues (in a field theory language there
should be no negative squared
masses).

If we consider the case where $y=0$, the $S(n)$ symmetry is promoted to
an $O(n)$ symmetry (i.e. there is an accidental
symmetry). This  $O(n)$ symmetry is spontaneously broken by the
choice in eq. (\ref{SYM}).
Consequently the Hessian has zero eigenvalues (i.e. there are Goldstone
Bosons).

If $y$ is different from zero, the $O(n)$ symmetry is explicitly
broken and the Goldstone Bosons
acquire a mass squared proportional to $y$. A detail computation shows
that in the general case $y$ is  negative and the Goldstone Bosons
acquire a negative mass squared (i.e. the Hessian has negative
eigenvalues).

This instability implies that the proposed stationary point
is not a local  minimum and one has to look for a minimum which will be
no more invariant under the  $S_n$ group.
Such a minimum can be constructed \cite{MPV,PB2}; we have to introduce an
infinite
sequence of steps breakings the
$S(n)$ symmetry. After a rather long computation one finally finds a
matrix $Q$ to which one can
associate a function $q(u)$, where $u$ belongs to the interval $0-1$.

There is a simple physical interpretation of this symmetry breaking.
The presence of frustration
implies that for each realization of the couplings there are
different equilibrium states with
different local magnetisation (i.e. different {\sl vacua}). We
indicate by $w^\al$ the probability
of finding the system in the state labelled by $\al$ and by $m^\al_i$
the local magnetization (i.e.
the expectation value of $\s_i$ in the state $\al$). The overlap among
different states  may be defined as
\be
q^{\al \g}= {\sum_{i=1,N} m^\al_i m^\g_i \over N}.
\ee

After some computations one finds that
\be
\ba {\sum_{\al \  \g} w^\al w^\g f(q^{\al \g})} \equiv  \int dq P(q)
f(q) = \int_0^1 du f(q(u)).
\ee
The probability of having two states with given overlap $q$ is thus
controlled by the function $q(u)$.

This  approach is able to explain in a qualitative and sometimes
a quantitative way
many of the properties of real spin glass. When the replica  symmetry
is broken, a change in the external magnetic field changes the structure of
equilibrium states  and one must be quite carefully in the definition of the
susceptibility. One finds that there   are  two susceptibilities:
\begin{itemize}
\item
The linear response susceptibility which quantifies the response
to   a small variation of the
magnetic field measured on a short time scale such that no global
rearrangement of spins is
possible. This susceptibility is given by $\chi_{LR} =\beta (1-
q_{EA}$.
\item
The thermodynamic susceptibility which quantifies the response to  a
small variation of the
magnetic field on a very long time scale such that  global rearrangements
of spins are
possible. This susceptibility is larger\footnote
{One finds that $q_{EA}= \max (q(u))$).}
 and it is given
by $\chi_{eq} = \beta (1-\int du q(x))$.
\end{itemize}

The difference between these two susceptibilities is one of the most
characteristic phenomena experimentally observed in spin glasses and it
is well explained by this theory\cite{Bi86}.

 At this stage of the theory, where fluctuations are neglected, is
not clear which of the theoretical predictions does survive when the
effect of the fluctuations is included. In order to have more precise
theoretical prediction for real three dimensional systems it is
necessary to go beyond  the mean  field approximation. This  has been
the subject of intensive studies. Unfortunately the computation of the
corrections to the mean field theory are rather  involved. The
correlation functions at zero loops are rather complicated \cite {DK1}  and the
one
loop  corrections to the correlation functions have not yet fully
computed.

Numerical simulations in $4$ dimensions \cite{PR,CPR} strongly
suggest the correctness of the broken
replica picture. No anomaly is observed. Numerical simulations in $3$
have not produced a compleately clear picture, although some progresses
have recently been done \cite{CPS,MAPARI3}.

When we will master the
loops corrections, we will be hopefully able to set up a renormalization
group study of the properties  of the system in the low temperature
region and in particular to compute the value of the
lowest critical  dimension, i.e.
the dimension where the structure of replica symmetry breaking
scheme is no more consistent and the predictions of the mean field theory
do not apply anymore\footnote
{For the $O(n)$ symmetry the lowest critical dimension is $2$, for the
$Z_2$ symmetry it is $1$.}.

 \section{Other Gauge Groups}

There is no need to consider only the $Z_2$ group. The
generalisations to other groups are very
interesting. In particular the case of the $U(1)$ group is relevant
for the study of irreversibility
in high $T_c$ superconductors \cite{GG3}. Here we consider a different gauge
group,  non-Abelian, for
which new results can be found.

In this model each spin can take $p$ values\cite{POTT1}. The gauge
field $U_{i,k}$ is an element of the
permutation group of $p$ elements\cite{POTT2}. The Hamiltonian is
\be
H_U[\s]= \sum_{i,k} \delta_{\s_i U_{i,k} \s_k}\ .
\ee
The gauge group is the group of permutation of $p$ objects ($S(p)$).
If $p=2$ we recover the
previous case ($Z_2=S(2)$).

 Due to the non Abelian nature of the gauge
group there is no global $S(p)$ symmetry at fixed $U$. As an effect of this
decrease in the symmetry  the effective free energy contains
an extra cubic term
\be
W(Q)= \tau \Tr(Q^2) +g_3 \Tr(Q^3)  +z \sum_{ab}Q^3_{ab} +O(Q^4),
\ee
which is forbidden in the Abelian case at zero magnetic field.

The value of $z$ is crucial. If $z$ is small with respect to $g_3$
($z=0$ at  $p=2$) we find that the value
of $Q$ at the saddle point is a continuous function of the
temperature. For large values of $z$ (in
mean field we need $z>4$) we fins that the value of $Q$ jumps
discontinuously at the
phase transition $T_c$.

A dynamical computation shows that in these conditions
there are metastable states
above the critical temperature, which correspond to vacua having high
free energy of the true
ground states.
The system may be trapped in these false vacua, and it may remain
there for a quite long time \cite{CRIS,Cuku}.

If we consider only point independent minima of the effective
Hamiltonian we find that there is a
dynamical temperature ($T_D$) at which the characteristic time
($t_c$) of the system diverges\cite{KT1}.

Below $T_D$ the characteristic time diverges exponentially with the
size of the system.  This exponentially large time may be computed
as follows. We introduce an effective potential $V(q)$ defined as:
\be
- N V(q)= \int dP[\s] \ln \left( \int dP[\mu] \delta(q-q_{\s
\mu}) \right),
\ee
where
\be
q_{\s \mu}={\sum_{i=1,N} \s_i \mu_i \over N}
\ee
 and for simplicity we use the short notation
\be
\int dP[\s] \equiv \sum_\s \exp(-\beta H_U[\s]).
\ee

Using
the techniques of \cite{FPV1,KUPAVI} this potential can be computed by
considering the effect of coupling $R=1+\eps$ replicas in the limit
$\eps \to 0$. It turns out that this potential has always a minimum
at $q=0$. For $T<T_D$ we finds that this potential acquires a new
minimum at $q=q_m$, the two minima being obviously separated by a
maximum at $q=q_M$.

If we assume that the dynamics may be approximated by the motion of
the system in this potential, we find that
 \be
t_c \sim \exp (N \delta(T)),
\ee
where the function $\delta(T)$ is different from zero below $T_D$ and
is given by
\be
\delta(T)= V(q_M)-V(q_m),
\ee
i.e. by the free energy barrier that the system has to cross.

A characteristic time which increases exponentially with the size of
the system cannot be the correct answer in a system with a short
range Hamiltonian. It is well known that in this case metastable states
do not exist: false vacua  do decay by thermodynamic tunnelling and the
characteristic time can be computed by evaluating the free energy of
the critical droplet (or instantons in a field theoretical language).

If we use the thecnique of \cite{FPV2} to study the instantons of these
theories  we
find out that  the characteristic time is finite also below $T_D$: \be t_c
\sim \exp( \Delta(T)), \ee where the function $\Delta(T)$ diverges at the
critical temperature  as
\be
\Delta(T) \propto (T-T_c)^{-(D-1)} \label{INS}
\ee

These results can be qualitatively easily explained. The situation is
quite similar to the usual
enucleation theory in which we have to estimate the size of the
critical droplets. The time for
escaping from the false vacuum is given by the exponential of the
instanton action (or the
exponential of beta times the free energy barrier). An instanton of
radius $R$ has two contributions
to its action
\be
A(R) = -\tau R^D + \Sigma R^{(D-1)}
\ee
The first is a bulk contribution, which vanishes at the critical
temperature ($\tau \propto T-T_c$)
and the second is an surface term (the free energy of the interface).
The radius for which $A(R)$ is
maximum diverges as the inverse of $\tau$ and the free energy excess
at this point is proportional to
$(T-T_c)^{-(D-1)}$.

Near the critical temperature the evolution of the systems is
dominated by the rearrangement of spins
in large domain, whose radius $R$ diverges at the critical
temperature.  Larger and larger barriers
must be crossed when we go near $T_c$ and the characteristic times
diverges at this temperature.

 In the $Z_2$ case discussed in the previous section there are no metastable
states
already in the mean field approximation \cite{Cuku2,FM}. In short range
 models
the characteristic time is divergent at the transition, but it divergent  as a
 power
law, i.e. as $(T-T_c)^{-\la}$, where $\la=2$ in the mean field approximation.
 The
new feature of the non Abelian  models
is the exponential divergence of
the characteristic time at the  phase transition. This divergence
makes extremely hard to
thermalize the systems also at temperatures not so close to $T_c$.

 \section{Toward Real Glasses}

 In spin glasses the Hamiltonian is random as an  effect of quenched
random
disorder. In real glasses the  Hamiltonian is not random and the
quenched disorder is dynamically
generated at low temperature. We  can therefore ask how much of the
qualitative and quantitative
results which have obtained in spin  glasses may be transferred to
real glasses \cite{KT2}$^-$\cite{KT3}.

In order to understand this point  one can study  models in
which the Hamiltonian does not
contain quenched disorder  and to compare the results with those
coming
of random Hamiltonian\cite{MIGLIO}$^-$\cite{BM}.

In general our strategy is the following. We want to study the
properties
of a given Hamiltonian
$H_G$ which is not random. We consider a class of Hamiltonians
$H_R$, of which $H_G$ is a
particular case. We choose the class $H_R$ in such a way that the
statistical properties of $H_G$
and  that  of a  generic Hamiltonian in $H_R$ are as similar as
possible.
In the best case we can
obtain that the two  corresponding free energies coincide in the
high
temperature expansion. In
general the behaviour of  the system can be  controlled better in the
high
temperature phase.

After having constructed $H_R$ in an appropriate way, we can suppose
that  the statistical
properties   of $H_G$ and $H_R$ are the same or, if they are
different,
we can construct a
perturbative expansion  which compute this difference. It is clear
that
this approach may be
successful in the high  temperature region
(more or less by
construction) and it may also
reproduce the behaviour of the system in the  glassy region, included the
dynamic  and
static transitions. However
it is cannot certainly reproduce the  possible existence of an
ordered  {\sl crystal} phase.

The proposed strategy for approximating a given Hamiltonian with a
random Hamiltonian works very
well in the many cases, at least in the framework of the mean field
approximation\cite{MPR}$^-$\cite{MPR2}. Real glasses have a qualitative
behaviour that is quite  similar to the one described in
the previous section\footnote
{Theoretical arguments can be done
which point in this direction \cite{CKPR}.}. It
is therefore tempting to suppose that also  real glasses belongs to the
same universality classes of system with a random Hamiltonian. If this
conjecture is correct
equation  (\ref{INS}) should apply also to real glasses.

It well known that  glasses show an extremely large increase in
the value of the viscosity
(which is proportional to the characteristic time) near the glass
transition. This increase of fifteen
order of magnitude has been measured in many materials. In the
case of fragile glasses this  increase has  been fitted by the
phenomenological Vogel-Fulcher law \cite{VF}:
\be
\eta \propto \exp (C (T-T_c)^{-1}).
\ee
The prediction of the replica approach for a three dimensional
systems is
\be
\eta \propto \exp (C (T-T_c)^{-2}). \label{NEW}
\ee
 Unfortunately the data are not sensitive enough to make a clear cut
choice among the two proposed
formulae, although the data are fitted slightly better by eq. (\ref{NEW}) than
by the traditional Vogel-Fulcher law.  More careful  studies of this point are
needed
\footnote{A comparison with  other theories\cite{A1,A2} of the glass
transition would make this talk too long.}.

It is quite amazing that starting from the study of some peculiar
forms of gauge theory on the
lattice we finally end up with a partial theoretical understanding
of the phenomenological Vogel-Fulcher law, which
have been already proposed when Oscar Klein was young.

The possibility of applying these
geometrical ideas (which originate from particle physics) to such
far away systems as spin glasses
and real glasses, shows the substantial unity of theoretical physics, a
field to which Oscar Klein
gave so many and important contributions.

 \section{Acknowledgements} It is a pleasure for me to thank for many
useful discussions and the
very pleasant collaboration on these  problems L. Cugliandolo, S.
Franz, M. Giura, J.  Kurchan, E. Marinari, L. Pietronero, F. Ritort
and M. Virasoro.


\section{References}
 \begin{thebibliography}{99}

\bi{EA} S. F. Edwards and P. W. Anderson, J. Phys. F {\bf 5}, 965 (1975).

\bibitem{Bi86} K. Binder and A. P. Young,
Rev. Mod. Phys. {\bf 58} (1986) 801.

\bi{TOU} G. Toulouse, Comm. on Phys. {\bf 2}, 115 (1977).

\bi{MPRI}E. Marinari, C. Parrinello and R. Ricci, Nucl. Phys. {\bf B362}, 487
(1991).

 \bibitem{MPV}  M.~M\'ezard, G.~Parisi and  M.~A.~Virasoro,  {\em
Spin Glass Theory  and Beyond}, World Scientific, (Singapore 1987).

\bibitem{PB2} G.~Parisi, {\sl Field Theory, Disorder and
Simulations}, World Scientific, (Singapore 1992).

\bibitem {DK1} C. De Dominicis and I. Kondor, Europhys. Lett. {\bf 2}, 617
(1986).

\bibitem {PR} G. Parisi and F. Ritort, J. Phys. {\bf A 26},  6711 (1993).

\bibitem {CPR} J. C. Ciria, G. Parisi and F Ritort, J. Phys. {\bf A 26}
6730 (1993).

\bibitem {CPS} S. Caracciolo, G. Parisi, S. Paternello and N. Sourlas,
Europhys. Lett. {\bf 11}, 783 (1990)

\bibitem{MAPARI3} E. Marinari, G. Parisi, F. Ritort, J. Phys. {\bf A 27}
2687 (1994) and in preparation.

\bibitem {GG3} J.D. Reger, T.A. Tokuyasu, A.P. Young and M.P.A. Fisher, Phys.
Rev. {\bf B 44}, 7147 (1991).

\bi{POTT1} D Elderfiedl and D. Sherrington, J. Phys. {\bf C 16}, L971 (1983)

\bi{POTT2}  D.J. Gross, I. Kanter and H. Sompolisky, Phys. Rev. Lett. {\bf 55},
304 (1985).

 \bi{CRIS} A. Crisanti, H. Horner and H.-J. Sommers,  Z. Phys. {\bf
B92}  257 (1993).

 \bi{Cuku} L. F. Cugliandolo and J.Kurchan, Phys. Rev. Lett. {\bf 71}, 1
(1993).

 \bi{KT1}T. R. Kirkpatrick and D. Thirumalai, Phys. Rev. Lett. {\bf 58},
2091 (1987).

\bibitem{FPV1} S. Franz, G. Parisi and M.A. Virasoro, J. Phys. {\bf A 18}, 72
(1992).

\bi{KUPAVI} J. Kurchan, G. Parisi and M.A. Virasoro, J. Physique {\bf 3}, 18
(1993).

\bibitem{FPV2} S. Franz, G. Parisi, M.A. Virasoro, {\sl Interfaces and Lower
Critical Dimension in a Spin Glass Model}, Nordita Preprint 94/21 S.

\bi{Cuku2} L. F. Cugliandolo and J. Kurchan; {\it On the out of Equilibrium
Relaxation of the Sherrington - Kirkpatrick  Model}, {cond-mat
9311016}, {\it  Weak-Ergodicity Breaking
in  Mean-Field Spin-Glass Models}, Fifth International
Workshop on Disordered Systems,  Andalo, 1994, cond-mat
9403040, to be published in Phil. Mag..

\bi{FM} S. Franz and M. M\'ezard; {\sl Off Equilibrium Glassy
Dynamics: a Simple Case}, LPTENS 93/39, {\it On Mean-Field Glassy Dynamics
out of Equilibrium}, { cond-mat 9403004}, LPTENS  94/05.

\bi{KT2}T. R. Kirkpatrick and D. Thirumalai, Phys. Rev. {\bf B36}, 5388 (1987).

\bi{Past} R. Palmer and D. Stein; in {\it Lectures in the science of
complexity}, D. L. Stein ed. (Santa Fe Institute, Addison-Wesley Pub. Co.,
1989).

\bi{KT3}T. R. Kirkpatrick, D. Thirumalai and P.G. Wolynes,  Phys. Rev. {\bf
A40}, 1045 (1989).

\bibitem{MIGLIO}G. Migliorini, {\em Sequenze Binarie in Debole
Autocorrelazione},  Tesi di Laurea, Universit\`a di Roma {\em Tor Vergata}
(Roma, March 1994).

\bibitem{MPR}E. Marinari, G. Parisi and F. Ritort,  {\sl Replica  Field
Theory for Deterministic Models: Binary Sequences  with Low Autocorrelation}
Hep-th/9405148,   to be published in J. Phys. {\bf A} (Math. Gen.); {\sl
Replica
Field Theory for Deterministic Models:  A non Random Spin-Glass Model  with
Glassy Behaviour}, Hep-th/9406074.

\bi{BM}J.-P. Bouchaud and M. M\'ezard; J. Physique I (France) {\bf
4} (1994) 1109.

\bi{Miri}  G. Migliorini and F. Ritort, Rome Preprint, cond-mat { 9407105},
to be published in J. Phys. {\bf A} (Math. Gen.).

\bi{Frhe} S. Franz and J. Hertz; Nordita Preprint, cond-mat {\ 9408079}.


\bi{ANA} G. Parisi, {\sl $D$-dimensional Arrays of Josephson Junctions, Spin
Glasses and  $q$-deformed Harmonic  Oscillators}, cond-mat preprint (1994).

\bibitem{MPR2} E. Marinari, G. Parisi and F. Ritort,  {\sl Fully Frustated
Ising Spin Model on the Hypercube is Glassy and Aging}, cond-mat
9410089 (1994) and in preparation.

\bibitem{CKPR} L. Cugliandolo, J. Kurchan, G. Parisi and F. Ritort, {\sl
Matrix Models as Solvable Glass Models}, cond-mat 9407086 (1994).

\bi{VF} H. Vogel, Phys. Z, {\bf 22}, 645 (1921); G.S. Fulcher, J. Am.
Ceram. Soc., {\bf 6}, 339 (1925).
\bi{A1} G. Adams and J. H. Gibbs, J. Chem. Phys. {\bf 28}, 139 (1965).
\bi{A2} G.S. Grest and M. H. Cohen, Phys. Rev. {\bf B 21} 4113 (1980).
 \end{thebibliography}
\end{document}